# Chiral Ordering Spin Associated Glass like State in SrRuO$_3$/SrIrO$_3$ Superlattice

Bin Pang[1], Lunyong Zhang[1,2*], Y.B Chen[3*], Jian Zhou[1], Shuhua Yao[1], Shantao Zhang[1], Yanfeng Chen[1]

1. National Laboratory of Solid State Microstructures & Department of Materials Science and Engineering, Nanjing University, Nanjing 210093, China
2. Max Planck POSTECH Center for Complex Phase Materials, Max Planck POSTECH/Korea Research Initiative (MPK), Gyeongbuk 376-73, Korea
3. National Laboratory of Solid State Microstructures & Department of Physics, Nanjing University, 210093 Nanjing, China

   * Corresponding authors:

   Lunyong Zhang allen.zhang.ly@gmail.com    and Y.B Chen ybchen@nju.edu.cn

**ABSTRACT**：Heterostructure interface provides a powerful platform to observe rich emergent phenomena, such as interfacial superconductivity, nontrivial topological surface state. Here SrRuO$_3$/SrIrO$_3$ superlattices were epitaxially synthesized. The magnetic and electrical properties of these superlattices were characterized. Broad cusps in the zero field cooling magnetization curves and near stable residual magnetization below the broad cusps, as well as two steps magnetization hysteresis loops are observed. The magnetization relaxes following a modified Stretched function model indicating coexistence of spin glass and ferromagnetic ordering in the superlattices. Topological Hall effect was demonstrated at low temperature and weakened with the increase of SrIrO$_3$ layer thickness. These results suggest that chiral ordering spin texture were generated at the interfaces due to the interfacial Dzyaloshinskii-Moriya (DM) interaction, which generates the spin glass behaviors. The present work demonstrates that SrIrO$_3$ can effectively induce interface DM interactions in heterostructures, it would pave light on the new research directions of strong spin orbit interaction oxides, from the viewpoints of both basic science and prospective spintronics devices applications.

**KEYWORDS:** Itinerant ferromagnet, spin orbit interaction, superlattices, spin glass, chiral ordering spin texture,

## I.  INTRODUCTION

Perovskite compounds with spin orbit interaction (SOI), for example the Sr$_{n+1}$Ir$_n$O$_{3n+1}$ oxides, was extensively studied in the recent years since SOI supplies another spin modulation mechanism beyond the magnetic field and Pauli exclusive interaction. SOI would delicately modify the crystal structure and electronic band structure, then lead to many emergent physical states, such as topological insulating states [1], unconventional superconductivity[2-3] and improper ferroelectricity[4]. It has been suggested in the $Sr_{n+1}Ir_nO_{3n+1}$ family, novel 1/2 effective total angular momentum state ($J_{eff}$=1/2) is formed due to the strong SOI accompany with strong electron correlation[5], which makes the family evolve from an canting antiferromagnetic Mott insulator ($Sr_2IrO_4$) to a paramagnetic semimetal ($SrIrO_3$, SIO) with the structural dimensionality from two dimensional to three dimensional. Up to now, the existing literatures have established the understanding of the basic properties such as magnetic structure, electronic band/orbit structure and transport behaviors[6-9] of the $Sr_{n+1}Ir_nO_{3n+1}$ family, especially $Sr_2IrO_4$ and $SrIrO_3$ owning to that the front is supposed to be a mother phase like the $La_2CuO_4$ to realize spin singlet d-wave high temperature superconductivity[10-11] and the latter is expected to be reconstructed into varied topological phases through breaking different structure symmetry of SIO[12-15]. Much effort, for example the Refs. 16-18, has been devoted to the superconductivity topic of $Sr_2IrO_4$, but it still has not been realized in experiment, depressing the prospection of it. The topological phase topic of SIO consequently attracted more and more attentions in recent years.

Most of the proposals for realizing topological phases in SIO are based on preparing heterostructures with SIO and other $ABO_3$ oxides[12-15], suggesting the expectation of novel physics in heterostructures of SIO. In principle, besides the proposed topological phases, the Dzyaloshinskii-Moriya interaction (DMI) is expected at the heterostructure interface of SIO due to the fulfilled DMI generating conditions of strong SOI accompanying with spatial inversion symmetry breaking[19-20], which would adjust spin state in the heterostructures, possible producing particular spin texture like Néel type Skyrmions[21]. We also propose realizing spin orbit torque (SOT) in the SIO heterostructures due to its strong SOI, with the same mechanism in magnet/heavy metal system[22-24]. SIO heterostructures is therefore a promising new research direction of strong SOI materials with giant interests. Recently, heterostructures of atomically thin SIO have been fabricated by the pulse laser deposition technique[25-26], achieving the first step towards the new direction. A close recent paper has set an good example, which reported that the magnetic easy axis reorientation in $La_{2/3}Sr_{1/3}MnO_3$/SIO superlattices is controlled by the SIO layer thickness [27]. The present paper epitaxially fabricated the superlattices combining varied thickness SIO layers with typical itinerant ferromagnetic SRO layers (fixed as 4 nm). Spin glass like state and SIO layer thickness dependent topological Hall effect were observed, demonstrated as interfacial DMI induced interface behaviors, highlighting the great potential of discovering novel physics in the superlattice constructed by strong SOI materials including $SrIrO_3$ and varied functional materials such as superconductors, magnets and topological insulators, suggesting a new research direction of iridates. Our results besides suggest that the interface DMI could be tuned by thickness of spin orbit interaction layer, as well as supply a practical method to tailor the properties of $SrRuO_3$ systems which has been often used in oxide spintronics.

## II. EXPERIEMTNAL SECTION

All the three groups of specimens were grown with five periods of 4nm-SRO/SIO on the chemical etching prepared flat (001)-SrTiO$_3$ substrates with Ti-rich termination[28], by a pulse laser deposition system (PLD, AdNaNo) at 2Hz repetition rate with substrate temperature at 800 ℃ and 75mTorr oxygen atmosphere. The SIO layer thicknesses are about 5nm, 10nm and 13nm respectively for the specimens (hereafter labeled as SL4/5, SL4/10 and SL4/13 as respectively), much thicker than the reported thickness for canting-AFM generation in SIO film, ~1.6nm (4 unit cells) [25]. The thicknesses of the alter layers ere measured by transmission electron microscopy (TEM, Tecnai F20). All the magnetic properties characterizations were carried out in a Quantum Design Magnetic Property Measurement System (MPMS3), and all the electron transport measurements were performed with the standard four-probe method in a 9T-Quantum Design Physical Property Measurement System (PPMS). The film surface morphology was imaged through an Asylum Cypher atomic force microscopy system. For the magnetization relaxation experiments, two kinds of field routes were adopted, one is the often adopted route that the samples were 0.1T field cooled to the target temperature from room temperature, and then removing the field following collecting the magnetization data (here noted as R$_{FC}$). Another route adopted is that the samples were first zero field cooled to the target temperature and then kept in an out plane magnetic field of 0.1T for 5 minutes, after that the field was removed the magnetization data were collected (noted as R$_{ZFC}$). Comparing with the former route, the latter one bears more powerful to distinguish the long range magnetic order state and the spin glass like state (see the support materials for details).

## III. RESULTS AND DISCUSSION

Figure 1a-1c present the surface morphologies of the samples, two dimensional growth model terraces were demonstrated, consistent with the distinct flat interfaces imaged by TEM and High resolution TEM (HRTEM), see figure 1d-1i, indicating no obviously atomic diffusion at the interfaces. X-ray diffraction (XRD) patterns demonstrate distinct superlattice satellite peaks (figure S2). Therefore it can be concluded that high quality superlattice samples were obtained, and no obvious structure differences amongst the three series samples except the layer thickness of SIO (see the patterns of XRD and HRTEM). Orthorhombic structures of the SRO and the SIO layers were initially confirmed by the XRD patterns (figure S2) and further by the scanning area electron diffraction (SAED) patterns shown in figure 1j and 1k, only slight splitting can be distinguished at the high index spots[29]. These characterizations substantiate that the SRO/SIO superlattices are epitaxially lattice constrained.

Figure 2a demonstrates the temperature-dependent magnetization under zero field cooling (ZFC) and 0.1Tesla field cooling (FC, perpendicular to the surface plane) conditions. Obviously for all the samples, sharply rising magnetization started at about 100K is revealed on both of the ZFC and FC curves. The Curie-Weiss law is not held at the magnetization rising regions (See figure S3), implying ferromagnetic transition is occurred there, which corresponds to the itinerant ferromagnetic transition of the SrRuO$_3$ layers. The Curie temperature $T_c$ indicated in the samples are lower than that of bulk SrRuO$_3$ about 50-60K since the SrRuO$_3$ layers thicknesses are at several nanometer scale[30] as well as some inevitable Ru vacancies[31].

At some lower temperatures, broad cusps split FC-ZFC are observed for all the samples. Denoting the broad cusp peak temperature as $T_f$, it decreases with the increase of magnetic field loaded for magnetization measurements (figure 2b and the inset figure). Moreover at the temperature regime corresponding to the broad cusps peak, prominent magnetization relaxation (figure 2c under the $R_{ZFC}$ route and figure 2d under the $R_{FC}$ route) and opened magnetization loops were observed in the samples (figure 3a). These features strongly suggest spin glass like state[32-34]. Here the magnetic nanoparticle associated superparamagnetic state can be excluded in our samples since it does not show hysteresis opened magnetization loop, even though it also demonstrates broad cusps spilt FC-ZFC. Below the left tail temperature of the broad cusp peak, near stable residual magnetization is observed (figure 2a and 2b), suggesting a FM background in the superlattices.

Quantitatively, the magnetization relaxation dynamics at the spin glass freezing temperature regime cannot be well fitted by the pure Stretched function $M(t) \sim \exp[-(t/\tau)^\alpha]$ which is usually held in simple spin glass systems [32] (see section 4 in supporting materials), but can be soundly tracked by a modified Stretched function model[35-36] with attaching one time independent baseline $M_0$, i.e. $M(t)=M_0+M_1\exp[-(t/\tau)^\alpha]$, see the solid lines in figure 2c and 2d. $\tau$ is the time constant, $M_0$ and $M_1$ are constants. $\alpha$ is a constant in a range from 0 to 1. Their fitting values are listed in the Table S1. The attached baseline term $M_0$ is interpreted as the FM contribution in the Gabay-Toulouse model which proves spin glass can coexist with long range FM ordering [37], right verifying our above statement that there is a FM background in the samples.

The magnetization loop (MB) measurements, as shown in figure 3a, also support the coexistence of spin glass and FM background in the samples. Two steps magnetization behaviors were shown in the loops for all the samples, which can be more clearly seen from the differential curves d$M$/d$B$ plotted in figure 3b, where two group peaks were exhibited. The peaks at low field corresponds to the coercive field of the main step magnetization, and those at high field indicate the second step magnetization. Similar two step magnetization loops were often observed in structure stacked by multiple magnetic layers, and was interpreted as a result of the stepwise switching of the different magnetic layers [38-39]. However, here the SIO layers should be paramagnetic since their thicknesses are much thicker than the critical thickness for generating canting-AFM in SIO films, 3 unit cells about 1.2nm[25]. Therefore it is not applicable to directly adopt the magnetic multilayer mechanism here; but nevertheless, it is natural to extend the idea of step switching of different magnetic layers to step switching of the spin glass texture and the FM background in the superlattices.

It was demonstrated that both of $T_c$ and $T_f$ are decreased with the increase of SIO layer thickness, see figure 2a and figure S5 (more clear showing $T_c$ variation). The depressed magnetization due to spin glass freezing defined as $(M_{FC}-M_{ZFC})/M_{FC}$ also decrease with the increase of SIO layer thickness, see the inset figure in figure 2a. Besides, the high coercive field switching on MB curves is gradually weaken as increasing the SIO layer thickness (figure 3). Therefore it is reasonable to propose that the observed spin glass like behavior is charged by an interfacial effect with the spin texture formed at the SRO/SIO interfaces, which induced the observed high coercive field step magnetization on the MB curves. The FM background left in the SRO body takes the responsibility of the near stable residual magnetization at the low temperature limit and the low coercive field magnetization on the

MB curves.

Particularly, we observed topological Hall effect (THE) in our samples, which further consolidates the interface effect conclusion. Figure 4a gives out the Hall traces of one SL 4/5 sample at different temperature regimes, good square shape opened hysteresis with the magnetization relevant anomalous Hall effect is distinct at the magnetic states ($T$<100K), just similar to that always seen in bulk $SrRuO_3$ [40]. Specially, peaks of $R_{xy}$ as highlighted by the arrows are emerged on the 10K Hall trace, which was never seen in $SrRuO_3$ system. Same emergent $R_{xy}$ enhancing peak also was seen for the SL 4/10 samples (pointed out by arrows in figure 4b, where the ordinary Hall effect components have been deducted from the original Hall resistances. Original $R_{xy}$ curves of SL 4/10 and SL 4/13 are shown in the figure S7). Moreover, the magnetic field of the peaks appearing are right consistent with the field of the second step magnetization on the MB curves, means that they are originated from same spin structure (figure 4d). Similar emergent behavior has been before reported in systems such as the itinerant-electron magnet $MnSi$[41-42], the centrosymmetric EuO thin films [43] and the magnetic/non-magnetic topological insulator heterostructures, $Cr_x(Bi_{1-y}Sb_y)_{2-x}Te_3/(Bi_{1-y}Sb_y)_2Te_3$ [44], and is called THE. The mechanism is assigned to the Berry phase in real space constructed by the chiral ordering spin texture[45-46]. The chiral ordering spin moments would be parallel aligned with increasing magnetic field, so depressing the Berry phase, which induces the decrease side of the THE peak. Further we note that the THE signal is weaken in the SL 4/10 samples and disappeared in the SL 4/13 samples (figure 4b), illustrating that the SIO layer plays a basic role in producing the THE associated spin texture which would be gradually destructed as the SIO layer thickness increasing.

Principally, chiral ordering spin configuration would be arose due to the Dzyaloshinskii-Moriya interaction (DMI)[19-20], i.e. the antisymmetric exchange coupling between two neighboring magnetic spins. It was discovered in recent years that noticeable interfacial DMI can be induced in multilayers constructed by ultrathin FM layer and nonmagnetic heavy metal (HM), owning to the large SOI (supplied by HM) with broken inversion symmetry at the interface[47-48]. That is right coincidence to the investigated system in the present work, where SRO is ferromagnetic and SIO owns strong SOI. Consequently, we can conclude that the interfacial DMI arises chiral ordering spin texture at the SRO/SIO interfaces, and which induces the THE and the spin glass behaviors. The mechanism to generate spin glass like behaviors by chiral spin configurations has been proposed by Kawamura[49-50], and has been well validated by Monte Carlo simulations[50] as well as was believed realizing in canonical spin glass magnet AuFe[51-52]. Its basic point can be manifested as that chiral spin configurations introduce spin frustration which is the essential condition to arise spin glass like state[32].

The demonstrated SIO layer thickness dependence of the THE as well as other magnetization features shown before highlights the interfacial DMI strength is tuned by the SIO layer thickness, here suggesting an inverse relationship between them. It is unfortunately, to our best knowledge, so far no a specific theory was established to profile the physical origination of the thickness dependence of interfacial DMI strength. We state a plausible interpretation based on the essential premises for generating interfacial DMI that strong SOI and broken spatial inversion symmetry. It is well known that the Rashba SOI arises also due to broken spatial inversion symmetry, so similarly often is significant in heterostructures. If proposing the interfacial DMI in the present system is a result of the Rashba SOI (Rashba SOI strength is proportional to atomic SOI and the potential asymmetry in the direction perpendicular to the interface, and Rashba SOI would magnify the effect of atomic SOI in heterostructures), the observed SIO thickness dependence of DMI would be understood since Rashba SOI strength is decreased with increasing the thickness of the SOI material layer constructing the heterostructures[53-54], due to the decreased potential asymmetry across the interface as the SOI layer being thicker (that is a result of quantum size effect[55]).

Magnetoresistance, defined as MR=[$R$-$R$($B$=0)]/$R$($B$=0), could also indicate the information of spin dynamics. The MR shown in figure 4c, figure S8a) and S8b) further reveals the strong correlation between charge transport and the spin frozen process. Obvious hysteresis is displayed on the MR in the spin frozen regime. In contrast, no distinct hysteresis were shown in temperatures higher than the spin frozen regime. The hysteresis of MR is induced by the magnetization hysteresis. The chiral ordering spin textures associated with the THE and spin glass like state also was indicated by the MR. For the MR of SL 4/5 at 2K and 10K, the hysteresis positive peaks are formed through two steps with the field increase. The first step starting from zero field corresponds to the low coercive field magnetization process shown in the MB curves (figure 3a) and bears a lower variation rate of MR (figure 4d), which is always seen in the MR of magnets with hysteresis. It reveals the nonuniformly spin re-orientation process in ordinary ferromagnetic domain recovering under magnetic field reversion, reminding the near stable residual ferromagnetic section on the ZFC-MT curves. The second step highlighted by the arrows, featured as abrupt sharp increase of MR (also shown in figure 4d as the arrows labeling additional peaks in the d$MR$/d$B$ curve) occurs at field corresponding to that of the THE signal and the second step magnetization on MB curve, is achieved by recovering of the chiral ordering spin textures. Figure S8 exhibits that the first step is prevail in all the researched samples, but the second step is not observed in the samples SL 4/10 and SL 4/13 which is agree with that the chiral ordering spin texture is gradually weakened with increasing the SIO layer thickness as revealed in the front sections.

## IV. CONCLUSION

In summary, we have fabricated SrRuO$_3$/SrIrO$_3$ superlattices and probed their magnetic and electron transport behaviors. SrIrO$_3$ layer thickness dependent spin glass like behaviors and topological Hall effect were observed. The spin relaxation dynamics follows the modified Stretched function model valid for a system with coexistence of spin glass texture and ferromagnetic background. It is therefore concluded that chiral spin ordering structure are formed at the SrRuO$_3$/SrIrO$_3$ interfaces. Our discoveries experimentally display that interfacial Dzyaloshinskii-Moriya interaction can be induced at the heterostructures constructed with SrIrO$_3$ and other functional oxides, so generating versatile novel physics in the heterostructures. The present study enlarges the research area of the current extensively interested strong SOI 5d transition metal oxides, would trigger more further theoretical and experimental investigations in such systems, leading better understanding of the complex interactions of SOI and other energy degrees. Besides, the present work simultaneously suggests one way to tune the magnetic state of SrRuO$_3$ films, so paving impacts on the oxide base spintronic applications. It is unfortunate that we cannot quantitatively calibrate the interfacial Dzyaloshinskii-Moriya interaction strength and interface chiral ordering spin texture in our samples because the relevant particular facilities for their measurements such as the Brillouin light scattering technique[47] are not available by us.

Note added.—During the polishing of the present paper, we became aware of Ref.[56], which reported the topological Hall effect in bilayer SRO/SIO, and shown the SRO thickness depdendence of THE, as well as simultaneously expected that skyrmions are formed at the SRO side of the SRO/SIO interface based on a multilayer system Hamiltonian simulation. Our present work studied SRO/SIO superlattices, reports the SIO layer thickness dependence of THE, and also in depth demonstrated the magnetic behaviors and their SIO layer thickness dependence, reveals the correspondence between the THE and the magnetic structure.

## ACKNOLEDGEMENT

Thanks to Professor Di Wu and Dr Binbin Zhang of Nanjing University for the enlightening discussions. We would like to acknowledge the financial support from the National Natural Science Foundation of China (51402149, 51472112, 51032003, 11374140, 11374149, 10974083, 11004094，11134006, 11474150 and 11174127), and major State Basic Research Development Program of China (973 Program) (2015CB921203, 2014CB921103, 2015CB659400), and the Fundamental Research Funds for the Central Universities (20620140630)

**FIGURES**

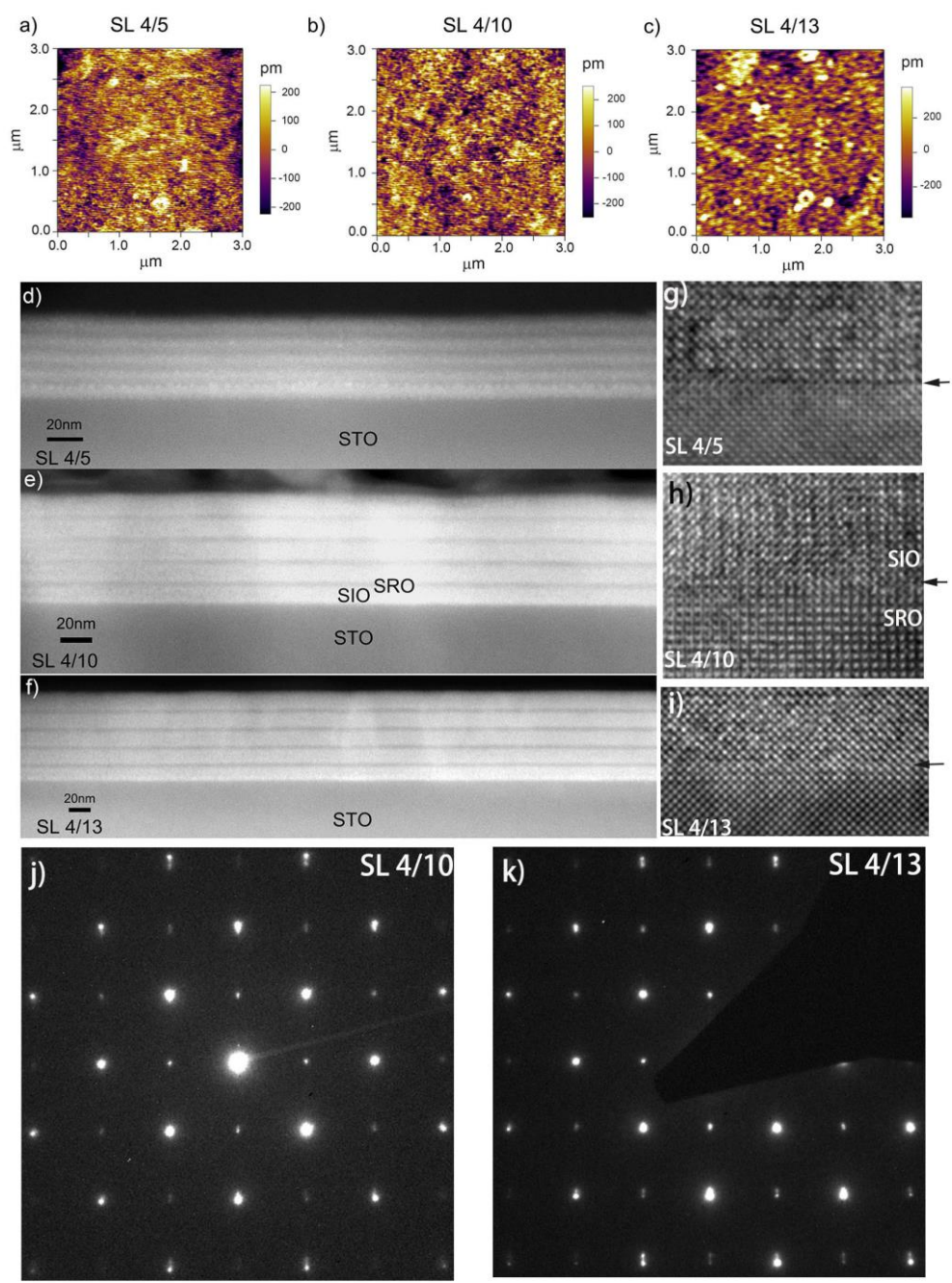

**Figure 1** a)-c) Atomic force microscopy images of the superlattices, showing the two dimensional growth model terrace surfaces; d)-f) and g)-i) TEM contrast images and HRTEM patterns of the surperlattice samples, respectively, illustrating the clear flat interfaces; j) and k) the SAED patterns of the SL4/10 and SL4/13 superlattices

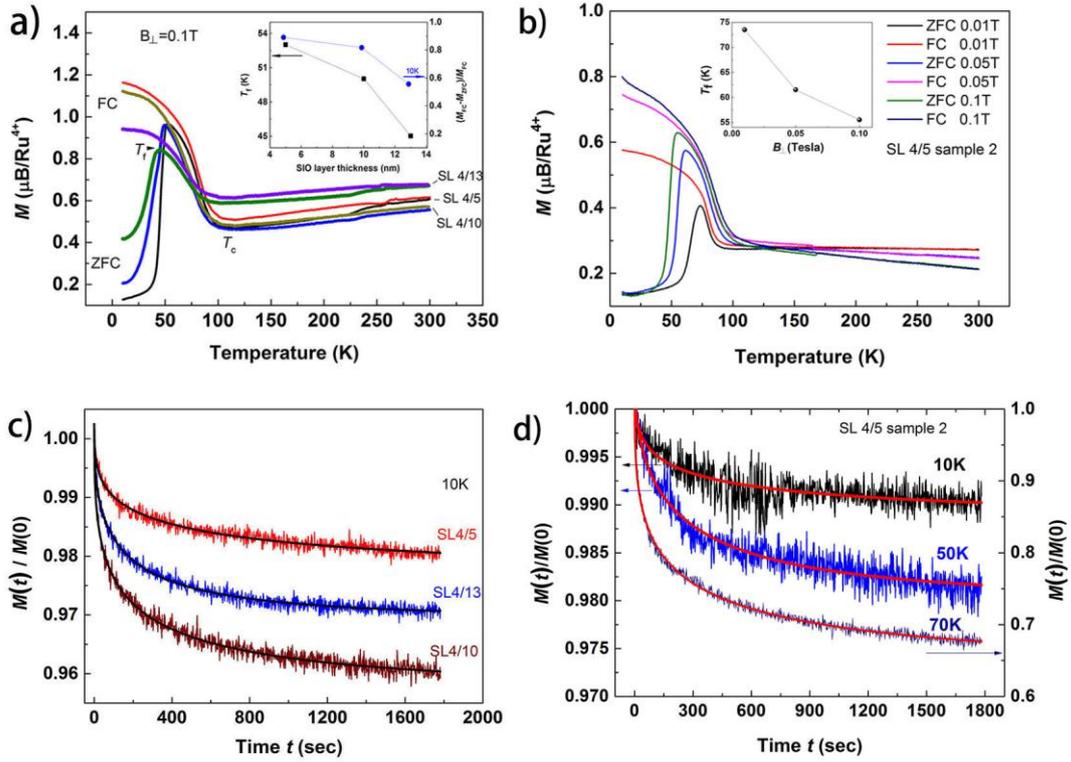

**Figure 2 Magnetic behaviors of the superlattices.** a) Temperature dependent magnetization curves (ZFC and FC) of the samples, inset showing the SIO layer thickness dependence of the spin glass freezing temperature $T_f$ and the depressed magnetization due to spin glass freezing defined as $(M_{FC}-M_{ZFC})/M_{FC}$. b) Temperature dependent magnetization curves of a SL 4/5 sample under difference magnetic field, inset showing the magnetic field dependence of the spin glass freezing temperature $T_f$; c) The magnetization relaxation curves of the samples at 10K under the $R_{ZFC}$ route; d) The magnetization relaxation curves of a SL 4/5 sample under the $R_{FC}$ route at varied temperatures. The relaxation data were fitted through the modified Stretched function model, shown as the solid curves.

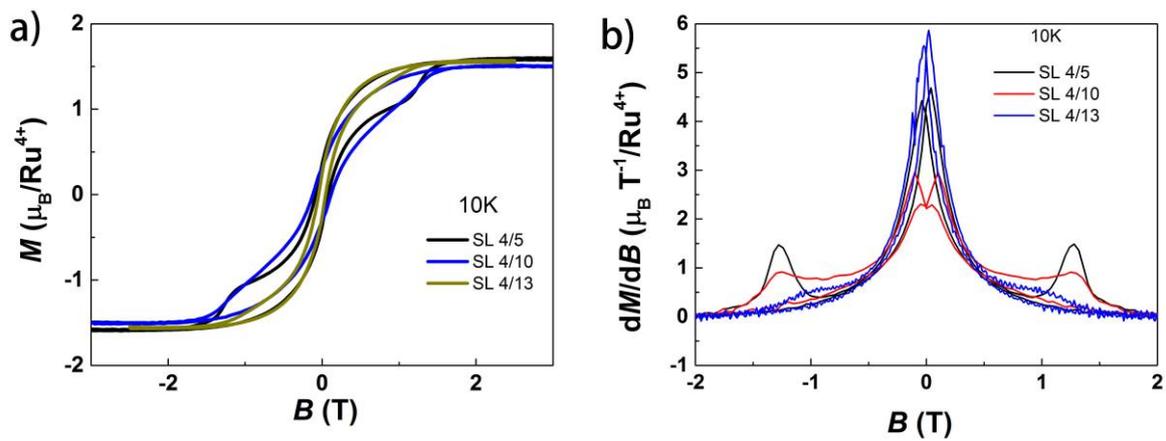

**Figure 3** a) **Magnetic hysteresis loops of the samples,** two step magnetization behaviors were demonstrated, suggesting that the spins undergo two switching processes.**; b)** Differential of the magnetization with respect to field.

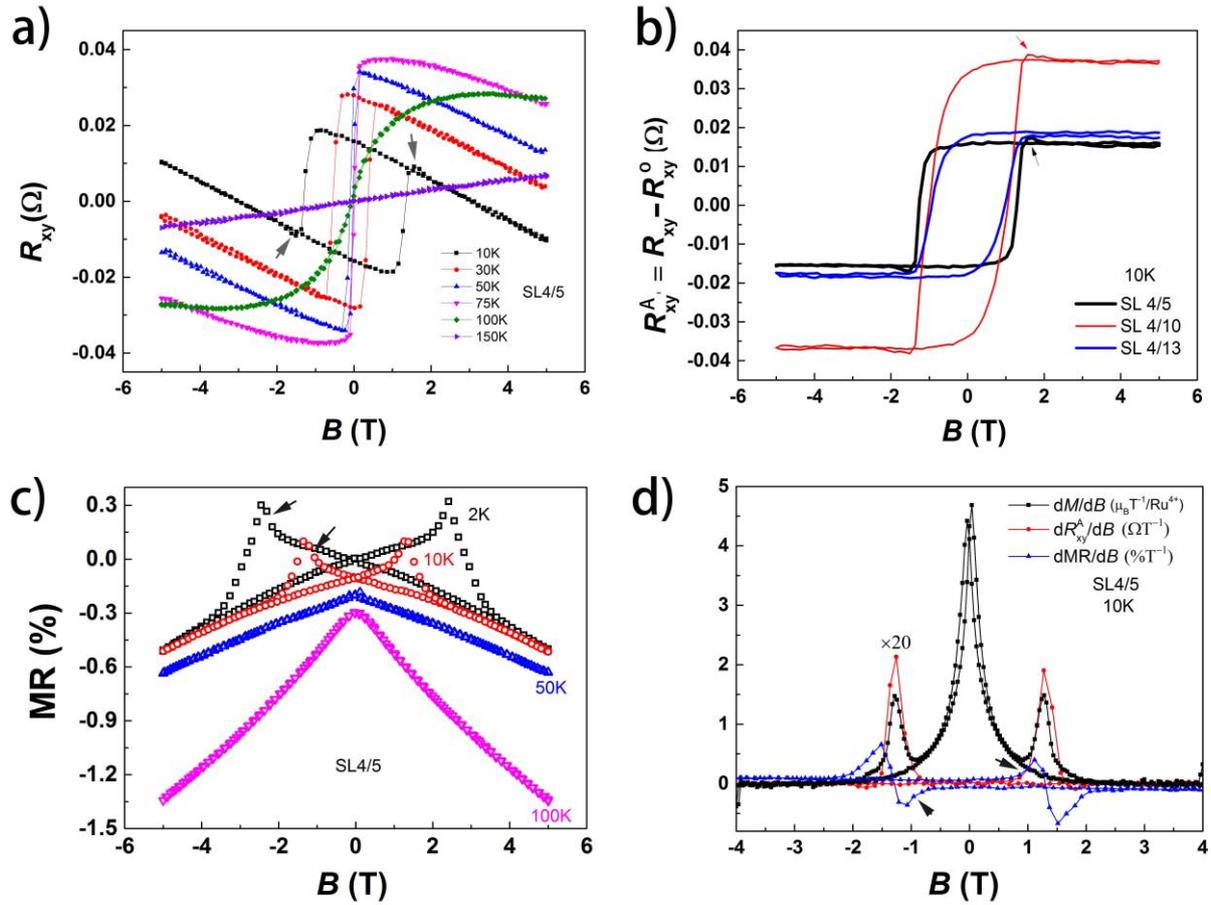

**Figure 4 Electron transport behaviors of the superlattices.** a) Hall resistances of SL4/5 at varied temperatuers, arrows highlighting the THE signal shown on the 10K trace; b) Hall resistances subtracted the ordinary Hall resistance components of the samples, at 10K; c) Magnetorsistances of SL 4/5, arrows highlighting the abrupt increased MR induced by the chiral ordering spin texture recovering on the 2K and the10K traces. The data of the 10K, 50K and 100K have been intentionally shifted -0.1%, -0.2% and -0.3% respectively for visual clarity. d) Differentials of the magnetization, Hall resistance and magnetoresistance of SL 4/5, demonstrating the field correspondence of their peaks.

*Supporting materials for*

# Chiral Ordering Spin Associated Glass like State in SrRuO$_3$/SrIrO$_3$ Superlattice

Bin Pang[1], Lunyong Zhang[1,2*], Y.B Chen[3*], Jian Zhou[1], Shuhua Yao[1], Shantao Zhang[1], Yanfeng Chen[1]

1. *National Laboratory of Solid State Microstructures & Department of Materials Science and Engineering, Nanjing University, Nanjing 210093, China*
2. *Max Planck POSTECH Center for Complex Phase Materials, Max Planck POSTECH/Korea Research Initiative (MPK), Gyeongbuk 376-73, Korea*
3. *National Laboratory of Solid State Microstructures & Department of Physics, Nanjing University, 210093 Nanjing, China*

*\* Corresponding authors:*
*Lunyong Zhang allen.zhang.ly@gmail.com   and Y.B Chen ybchen@nju.edu.cn*

## 1. R$_{ZFC}$ and R$_{FC}$ magnetization relaxation measurement routes

In the present work, the magnetization relaxation experiments were carried out through two kinds of field routes.

The R$_{FC}$ route: the samples were 0.1T field cooled to the target temperature from room temperature, and then removing the field following collecting the magnetization data. The magnetization relaxation data shown in figure 2c were measured under this route.

The R$_{ZFC}$ route: the samples were first zero field cooled to the target temperature and then kept in an out plane magnetic field of 0.1T for 5 minutes, after that the field was removed the magnetization data were collected. The magnetization relaxation data shown in figure 2d were measured under this route.

We believe the R$_{ZFC}$ route is more powerful to identify the spin glass like state than the R$_{FC}$ route. The reason is discussed as following. Without loss of generality, a two energy states system with spin glass state and long rang magnetic ordering state can be profiled as figure S1. If the system has a background state with FM, the system would not be relaxed no matter after the FC or ZFC because it has been located at the lowest energy background (figure S1 a). If the system has a background state with AFM, relaxation is not expected to be observed since it has been entered the lowest energy background (figure S1 b). Using R$_{FC}$, when the exchange interaction causing the AFM state is large, so a FC induced spin parallel state is highly unstable, the system will relaxed to the background AFM state instantaneously when the field for FC is removed, so it is not expected to observe distinct relaxation, but when the exchange interaction is relatively waken or due to unavoidable defect in practice, the relaxation would be not instantaneous, so distinct relaxation can be observed (figure S1 c).

If the system is as shown in figure S1 d) with a spin glass background state. Using $R_{ZFC}$, the system will directly enter the spin glass state during the ZFC, and then the small external field would arouse spin fluctuation and relaxation would be observed after the field is removed. Under $R_{FC}$, the system possible enter a metastable FM state which cannot relax when the barrier energy $E$ is larger than $k_BT$ ($T$ is temperature). When $E$ is smaller than $k_BT$, relaxation is expected.

Consequently, when relaxation is observed under the $R_{ZFC}$ route, long range magnetic ordering state is excluded, and spin glass state is most possible. In contrast under the $R_{FC}$ route, AFM state is also possible when observing magnetic relaxation. Whatever, magnetic relaxation is demonstrated in the measurements under both the $R_{ZFC}$ route and the $R_{FC}$ route, therefore we conclude a spin glass like state in our SRO/SIO superlattices.

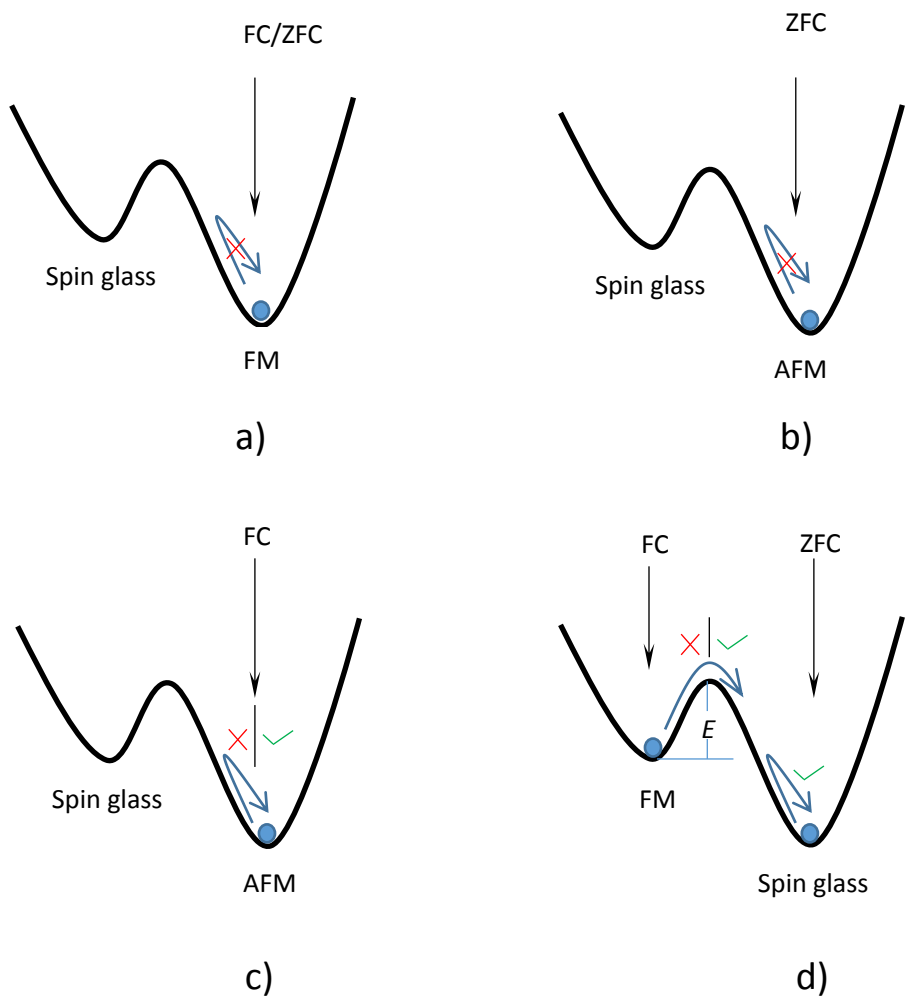

**Figure S1** Schematic energy diagram of a two energy states system with spin glass state and long rang magnetic ordering state. The cross symbol notes that the magnetic relaxation is not expected, and tick symbol notes that the magnetic relaxation is expected. $E$ is barrier energy.

## 2. XRD

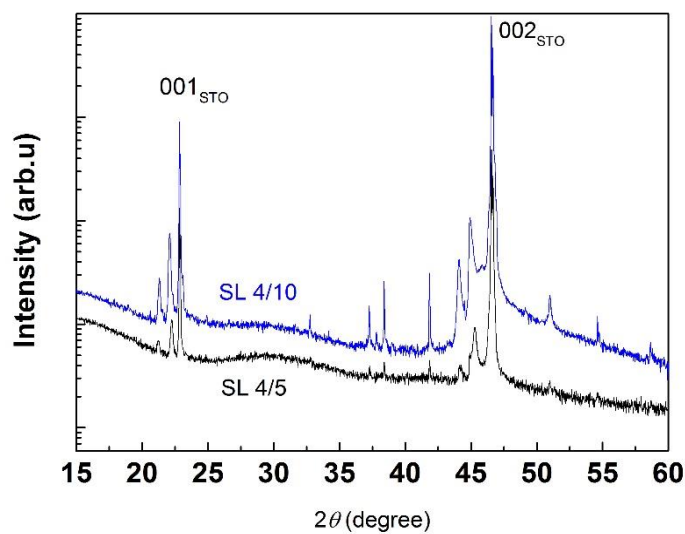

**Figure S2** Representative XRD patterns of the samples SL 4/5 and SL 4/10.

## 3. Curie-Weiss law representation of the MT

According to the Curie-Weiss law [1]

$$\chi = \frac{C}{T - T_0}$$     Equation S1

Figure S3 gives out the $1/\chi$-$T$ curves based on the ZFC data shown in figure 2a. No simple linear dependence below the $T_c \sim 100K$ was observed for the samples, meaning a magnetic transition there.

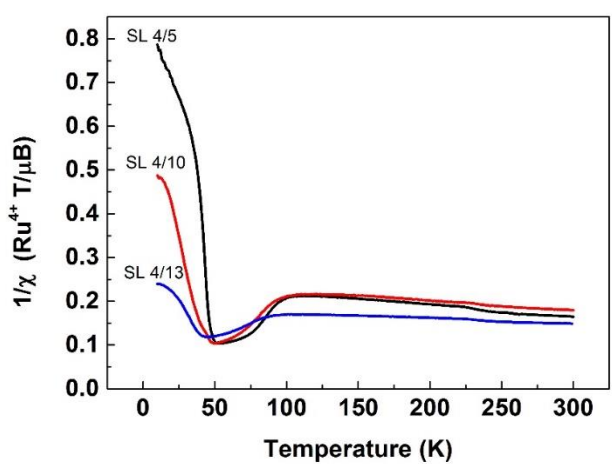

**Figure S3**   $1/\chi$-$T$ curves transformed from the ZFC data shown in figure 2a

## 4. Magnetization relaxation dynamics

The Stretched function usually adopted in description of relaxation phenomenon could be written as a general format as[2]

$$M(t) = M_0 \exp\left[-\left(\frac{t}{\tau}\right)^\alpha\right] \qquad \text{Equation S2}$$

So a linear dependent relation can be obtained as following

$$\ln\left[-\ln\left(\frac{M}{M_0}\right)\right] = \alpha \ln t - \alpha \ln \tau \qquad \text{Equation S3}$$

Figure S4 gives out the relaxation curves transformed according to Eq.S3. Obviously, no linear dependence could be assigned in the whole relaxation time.

The modified Stretched function is written as[3-4]

$$M(t) = M_0 + M_1 \exp\left[-\left(\frac{t}{\tau}\right)^\alpha\right] \qquad \text{Equation S4}$$

Which gives out the $M(0)=M_0+M_1$, so for normalized magnetization like we have adopted Eq.S4 is written into

$$\begin{aligned}
M'(t) &= \frac{M(t)}{M_0 + M_1} \\
&= \frac{M_0}{M_0 + M_1} + \frac{M_1}{M_0 + M_1} \exp\left[-\left(\frac{t}{\tau}\right)^\alpha\right] \qquad \text{Equation S5} \\
&= M_0' + M_1' \exp\left[-\left(\frac{t}{\tau}\right)^\alpha\right]
\end{aligned}$$

Table S1 gives out the fitted constants for the relaxation curves in figure 2c and figure 2d using Eq.S5. The fitted values of the exponential constant α for the 10K data of SL 4/5 are in agreement for the $R_{ZFC}$ and $R_{FC}$, indicating same structure basis in the two relaxation processes.

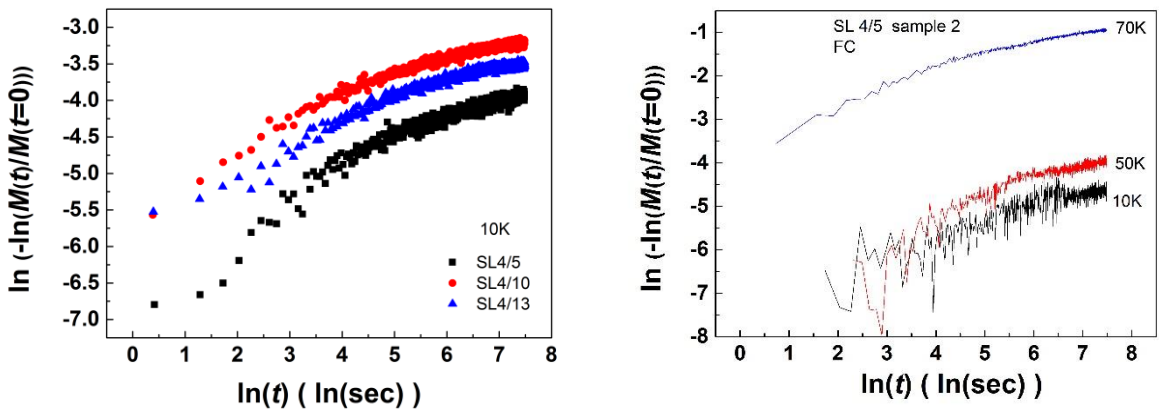

**Figure S4** Magnetization relaxation data transformed according to the Stretched function

**Table S1** Fitted constant values of magnetization relaxation according to the modified Stretched function model Eq.S5

| | $M'_0$ | $M'_1$ | $\tau$ | $\alpha$ |
|---|---|---|---|---|
| **Sample** | | 10K (figure 2c) | | |
| **SL4/5** | 0.976 | 0.024 | 361.922 | 0.359 |
| **SL4/10** | 0.957 | 0.043 | 155.419 | 0.391 |
| **SL4/13** | 0.967 | 0.033 | 151.903 | 0.491 |
| | | | | |
| **Temperature** | | SL 4/5 sample 2 (figure 2d) | | |
| **10K** | 0.988 | 0.015 | 216.068 | 0.350 |
| **50K** | 0.980 | 0.020 | 243.894 | 0.510 |
| **70K** | 0.636 | 0.364 | 202.851 | 0.376 |

## 5. Electronic transport

Figure S5 gives out the temperature dependent resistance of the samples, where resistance kinks indicating the itinerant ferromagnetic transition in SrRuO$_3$ layers are demonstrated. Inset more clearly shows the transition points $T_c$, it was shown $T_c$ is decreased with increasing the SrIrO$_3$ layer thickness. The low temperature resistance upturn may come from two contributions, one is the enhanced scattering coming from the spin glass freezing, and another one is owing to the weak localization of carriers. We noted the SIO layer thickness dependence of R': SL4/5<SL4/13<SL4/10. It could be explained as the SIO layer conductance component increases with increasing SIO layer thickness in the superlattice with parallel connection of SIO resistor and SRO resistor, so more and more feature coming from the R' of SIO layer would exhibited in the sample with thicker SIO layer. It is known the R' of SIO film is larger than that of SRO film since SRO is of better metallicity[5-6], for example see figure S6. As for why the R' (SL 4/13) is lower than R' (SL 4/10), we believe that comes from the stronger carrier weak localization in SL 4/10 due to inevitable and difficult controlled defects because weak localization caused resistance increase at low temperature will increase R'. It can be seen the metal transition temperature of SL 4/10 is higher than that of SL 4/13.

Figure S7 shows the Hall resistances of the samples SL 4/10 and SL 4/13. Hysteresis in the temperature regime below $T_c$ is obvious, which is similar to the Hall traces of SL 4/5 shown in the main text. Topological Hall signal could be slightly seen for the sample SL 4/10 at 10K, but no for the SL 4/13.

Figure S8 a) and b) present the magnetoresistances of the samples SL 4/10 and SL 4/13, respectively. Hysteresis was distinctly demonstrated at the temperatures at 2K and 10K. Figure S7 c) and d) give out the differential curves of the MR of the samples at 2K and 10K, the emergent variation of dMB/dB contributed by the recovering of the chiral ordering spin texture, pointed out by the arrows, is only distinct for the sample SL 4/5, indicating the weaken of the chiral ordering spin texture with increasing the SIO layer thickness.

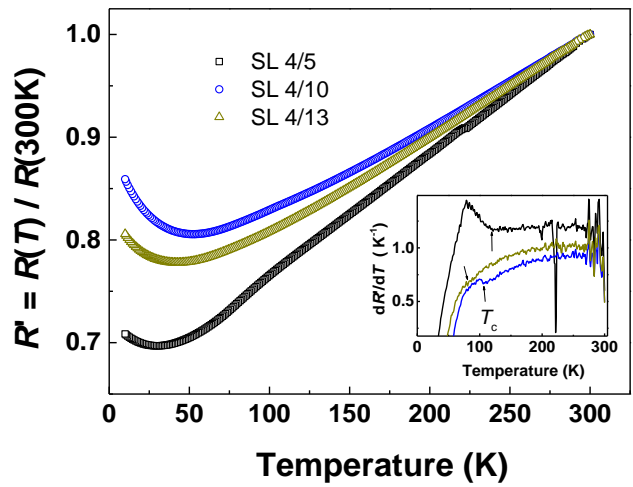

**Figure S5** Temperature dependent resistance of the samples. Inset showing the differential of resistance to determine the $T_c$ more distinctly.

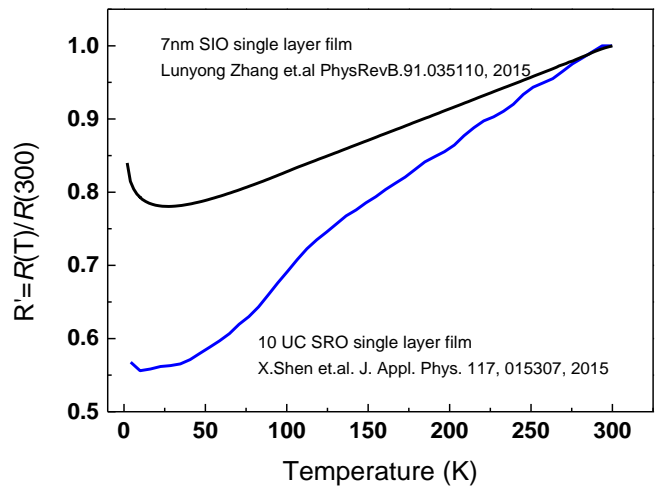

**Figure S6**  R' of SIO films and SRO films

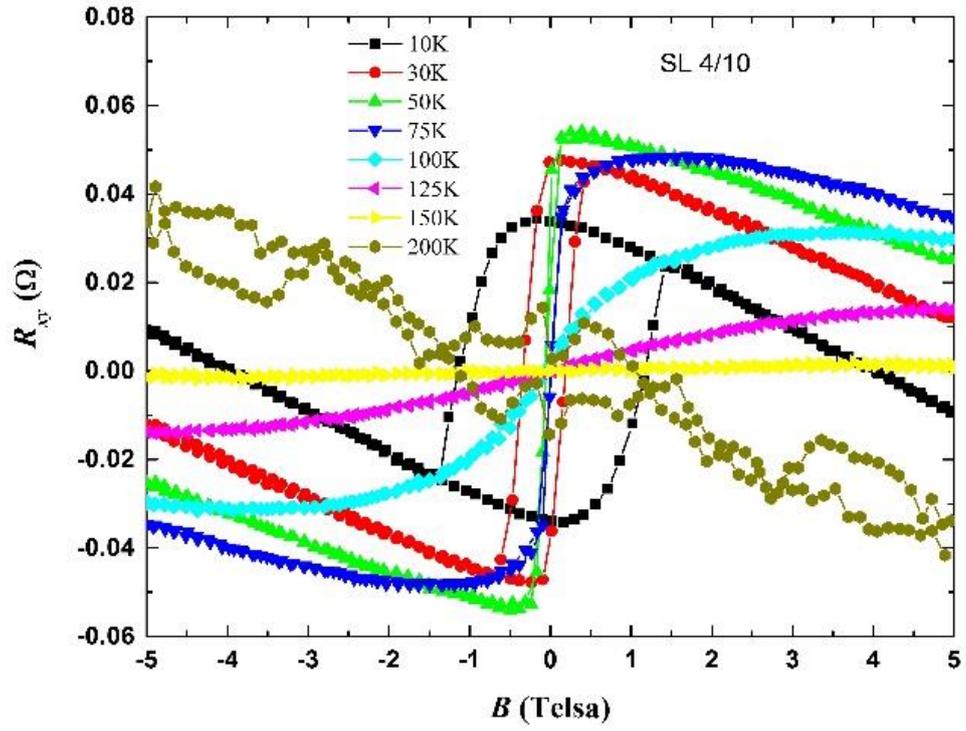

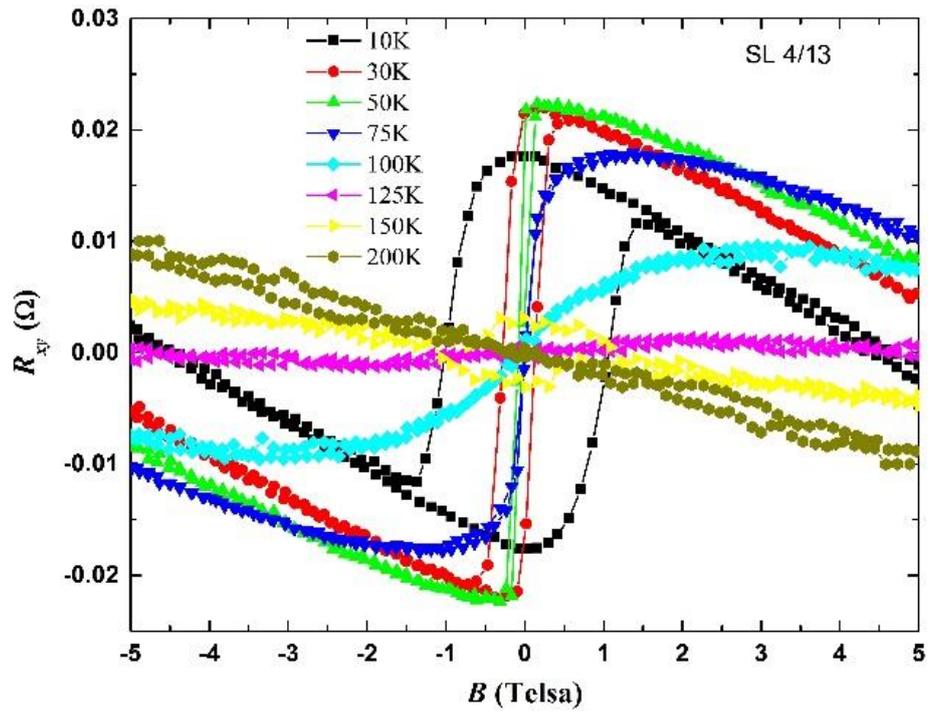

**Figure S7** Hall resistance traces of the samples SL 4/10 and SL 4/13

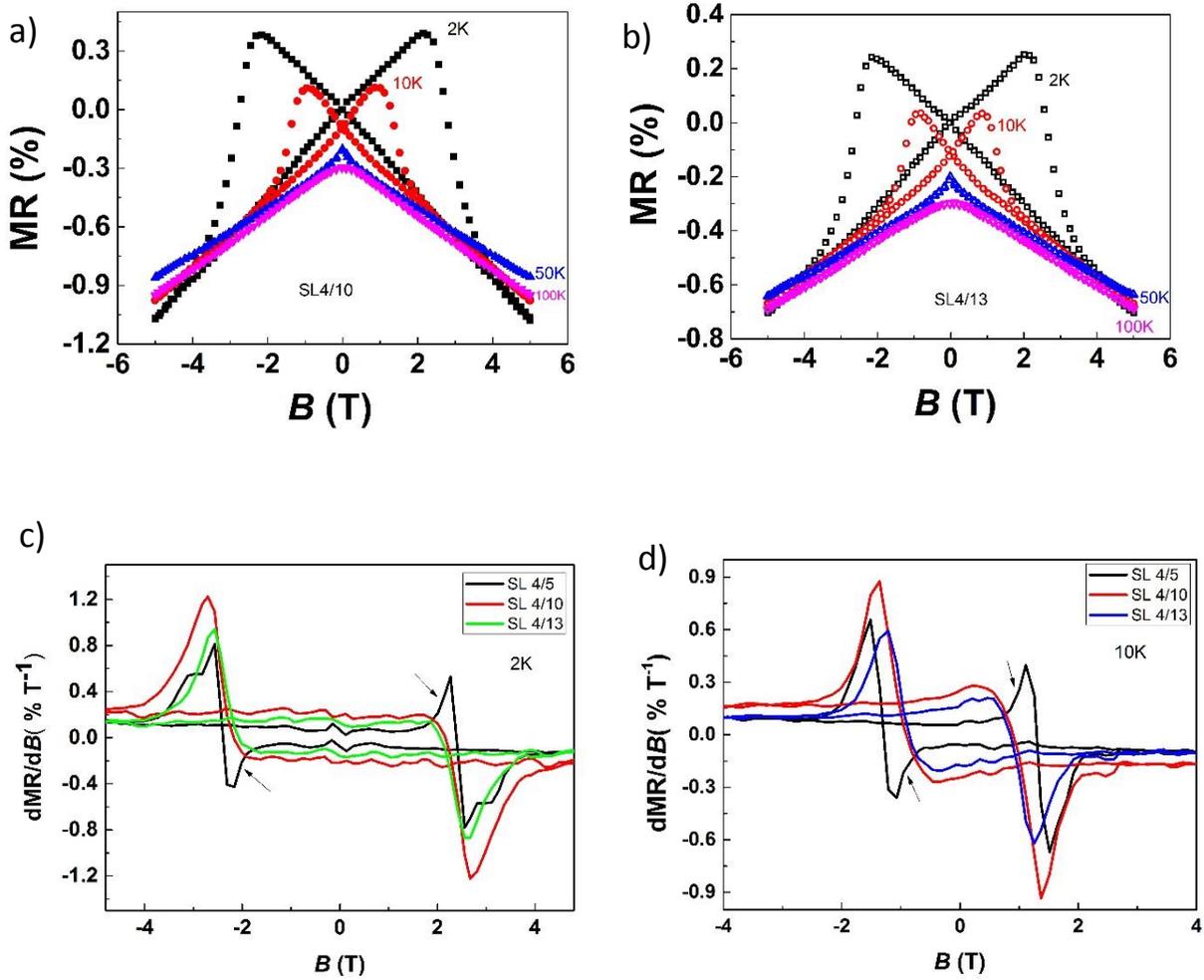

**Figure S8** magnetoresistances and their differentials of the samples